# Explainable Artificial Intelligence Techniques for Software Development Lifecycle: A Phase-specific Survey

Lakshit Arora
lakshit@google.com

Aman Raj
amanraj@google.com

Ankit Shetgaonkar
ankiit@google.com

Sanjay Surendranath Girija
sanjaysg@google.com

Dipen Pradhan
dipenp@google.com

Shashank Kapoor
shashankkapoor@google.com

Google

*Abstract*— Artificial Intelligence (AI) is rapidly expanding and integrating more into daily life to automate tasks, guide decision-making, and enhance efficiency. However, complex AI models, which make decisions without providing clear explanations (known as the "black-box problem"), currently restrict trust and widespread adoption of AI.

Explainable Artificial Intelligence (XAI) has emerged to address the black-box problem of making AI systems more interpretable and transparent so stakeholders can trust, verify, and act upon AI-based outcomes. Researchers have developed various techniques to foster XAI in the Software Development Lifecycle. However, there are gaps in applying XAI techniques in the Software Engineering phases. Literature review shows that 68% of XAI in Software Engineering research is focused on maintenance as opposed to 8% on software management and requirements.

In this paper, we present a comprehensive survey of the applications of XAI methods such as *concept-based explanations, Local Interpretable Model-agnostic Explanations (LIME), SHapley Additive exPlanations (SHAP)*, *rule extraction, attention mechanisms, counterfactual explanations,* and *example-based explanations* to the different phases of the Software Development Life Cycle (SDLC), including *requirements elicitation, design and development, testing and deployment, and evolution*.

To the best of our knowledge, this paper presents the first comprehensive survey of XAI techniques for every phase of the Software Development Life Cycle (SDLC). This survey aims to promote explainable AI in Software Engineering and facilitate the practical application of complex AI models in AI-driven software development.

## I. Introduction

Artificial Intelligence (AI)-aided software techniques supported by Large Language Models are rapidly transforming software development with increased productivity [1]. AI-aided Software Engineering is on the rise and is becoming vital to deliver reliable, valid, and maintainable software systems [2],[3]. However, trust and widespread adoption of AI are often hindered by the "black-box" problem, where complex AI models make decisions without providing transparent explanations for those decisions [5],[7]. Explainable Artificial Intelligence (XAI) has emerged as a non-functional requirement in AI systems to address this issue by improving the interpretability and transparency of AI systems, allowing stakeholders to trust, validate, and act upon AI-driven insights [5],[9]. With the growing attention to XAI, it has become challenging for practitioners and researchers to navigate and select appropriate XAI methods and tools for their specific applications [10].

Researchers have developed XAI methods for AI-aided software development. However, this area remains understudied [5],[11]. A blanket application of XAI to software engineering is insufficient. Different software engineering phases require tailored XAI techniques. For example, the literature reveals that inconsistent XAI evaluation methods in software engineering make it challenging to compare studies and XAI techniques across different software engineering phases [7],[12]. This disparity is evident in existing research, where studies indicate that 68% of XAI in Software Engineering research focused on the software maintenance phase versus 8% on the software management and requirements [7].

This paper aims to address the lack of XAI applications in the Software Development Lifecycle and recommend ways to apply it ethically and truthfully. We address this by finding XAI methods such as Local Interpretable Model-agnostic Explanations (LIME), SHapley Additive exPlanations (SHAP), and Rule Extraction, mixing and matching them to provide explainability in key Software Engineering phases: requirements elicitation, design and development, testing and deployment, and evolution. The following research questions guide the paper:

- RQ1: What are the key explainability challenges that AI introduces in software engineering?
- RQ2: How can tailored XAI techniques for each software engineering phase enhance the explainability of AI-aided Software Engineering?
- RQ3: What are the limitations of existing XAI techniques in Software Engineering?

A mixed-methods approach was employed to address these research questions, combining a systematic literature review (SLR) with a narrative review of relevant literature. The SLR analyzed existing XAI in Software Engineering studies to identify key themes, evaluate XAI methods, and identify research gaps. Searches were conducted in IEEE Xplore, ACM Digital Library, Science Direct, Wiley, Google Scholar, and Scopus using keywords such as "XAI", "Explainable Artificial

Intelligence", "Software Engineering", "AI-aided development", "trust", "transparency", and "ethical AI". The Inclusion criteria focused on peer-reviewed articles published within the last 6 years that addressed XAI in software engineering. A narrative literature review complemented the SLR to explore broader perspectives on XAI and Software Engineering, which assisted in providing XAI techniques for phase-specific Software Engineering. Through synthesizing SLR and narrative review results, this paper proposes the first comprehensive overview of XAI techniques tailored to each Software Development Life Cycle (SDLC) phase.

The paper is structured as follows: Section 2 reviews the literature on AI in Software Engineering, explainable AI, and XAI in Software Engineering. Section 3 discusses the proposed XAI techniques for each Software Engineering phase to improve the explainability of AI-aided software development. Section 4 presents the discussion, and Section 5 concludes the paper with recommendations and future research.

## II. LITERATURE REVIEW

### A. AI in Software Engineering

AI applications within the Software Engineering process are rapidly expanding and are considered significant, particularly with the use of Generative AI for tasks like code generation [21],[3]. Kokol (2024) conducted a comprehensive knowledge synthesis to assess the current status of published literature in AI in Software Engineering [11]. Martinez-Fernandz et al. conducted a comprehensive study on software engineering for AI-based systems, in which the authors did a systematic literature review of software engineering practices followed in AI-based systems [3]. Gorkem and Giray [22] conducted a study on Software Engineering for Machine Learning (ML) systems in which they outlined the misconception that sometimes arises between Software Engineering for ML which refers to Software Engineering approaches to developing ML or AI systems versus ML for Software Engineering which deals with the use of ML and AI in Software Engineering tasks [3]. This paper focuses on the latter, examining explainable AI (XAI) for Software Engineering to support the development of reliable, valid, and maintainable software systems. Specific AI usages in Software Engineering include:

- Requirements elicitation: Natural language processing for document analysis, chatbots for elicitation, data mining for user needs [11].
- Design and implementation: System architecture recommendations, user interface and experience generation, model selection, code generation, code completion, and bug detection [11].
- Testing and Verification: Test case generation, test prioritization, fault localization [25].
- Deployment and Monitoring: Performance evaluation, failure detection, bias monitoring [25].
- Maintenance: Bug prediction, refactoring recommendations, and change impact analysis [25].

Despite demonstrated effectiveness and efficiency, for example, correct code generation [21], AI adoption in Software Engineering is often marred by the "black-box" phenomenon, where AI outputs lack explanations understandable by software development stakeholders [27]. This leads to several limitations during the requirements elicitation phase *(e.g., ambiguity, lack of tacit knowledge extraction, bias, etc.)*, design phase *(lack of creativity, lack of trade-off analysis, lack of contextual awareness, etc.)*, code development phase *(e.g., lack of techniques for evaluating correctness, security vulnerabilities, maintainability, etc.),* and during the testing phase *(e.g., test oracle problem, test flakiness, scalability concerns, etc. [3],[28])*.

### B. Explainable AI (XAI)

Explainability is gaining traction within AI communities as a means to address the "black-box" problem, which stems from a lack of transparency regarding how AI models operate and arrive at their outputs [10],[29]. Explainable AI (XAI) techniques are designed to provide reasonable and understandable explanations on the complex decision-making processes of machine learning models [5], [30]. Vilone and Longo (2021) proposed a widely cited classification for XAI methods [30], based on the following properties:

- Stage of explainability: Refers to the period in the process of generating outputs when a model generates the explanation for the decision it provides. The authors discuss two stages called Ante-hoc and Post-hoc. Ante-hoc generates explanations for decisions from the beginning of the training data while aiming to achieve optimal performance, and Post-hoc, which provides explanations after the model has been trained and made predictions. Post-hoc can be either model-specific or model-agnostic. Model-agnostic methods apply to any model, while model-specific methods apply to specific models.
- Scope of explainability: Refers to the extent of an explanation produced by XAI methods. The scope can be local or global, with local explaining only an instance of inference to the user while global providing the entire inference of the model to the user.
- Input and output: Refers to the format of the input and the output that can be used by XAI techniques to explain the model decision. The XAI technique utilizes the XAI input to generate the explanation output. The XAI methods need to understand the same kind of data that the model itself uses. The most common forms of input explanations are images, text, and vectors, while output examples are numeric, rules, and visualizations.

In the following text, we present some of the most common XAI techniques and provide a brief description incorporating the stage, scope, and input/output properties. Figure 1 shows a brief summarization and classification of standard XAI techniques.

*1) Feature Attribution based techniques*: These techniques center around explaining predictions by assigning importance or relevance scores to the input features. They tell which features matter most. The most common feature attribution based techniques are:

- LIME (Local Interpretable Model-agnostic Explanations) [12],[31]: LIME is a post-hoc, model-agnostic technique that explains individual predictions. It works by approximating the complex model locally with a simpler, interpretable model *(e.g., linear model or decision tree)*. LIME generates new data points by perturbing the input features of the instance to be explained and then observing how the model's prediction changes. It then trains a weighted linear model on top of the instance, using the perturbed data and resulting predictions. This local linear model's weights are used as explanations, stating the relevance of each feature for that specific prediction. It can handle varying input types *(e.g., tabular, text, images)* and typically outputs feature importance as output *(e.g., numeric weights, or highlighted words/pixels)*. A practical example of this technique would be explaining why a particular image was classified as a "cat" by highlighting
the pixels that contributed most to the classification.
- SHAP (SHapley Additive exPlanations): Messalas, Andreas, et al. described SHAP as a post-hoc, model-agnostic XAI technique based on game theory [17]. This technique calculates Shapley values, which represent the average marginal contribution of each feature to the prediction expected over all possible feature combinations. SHAP values provide local explanations (explanation of one prediction) and global explanations (feature importance summary for the entire dataset). The input can be anything, e.g., the input of the original model. The output often features visualizable importance scores (numeric). An example of this technique would be explaining why a loan application was rejected by an AI model, showing the contribution of each factor (income, credit score, debt, etc.) to the rejection decision.
- Attention Mechanisms: These integrate into the model as part of the model architecture, rather than post-hoc. They are model-specific and occasionally used in deep learning models like Transformers for Natural Language Processing (NLP) and computer vision. Samek, Wojciech, et al. (2016) in their research [16] showcased how attention mechanisms provide a local explanation by highlighting the input areas (e.g., words in a sentence, regions in an image) the model is paying attention to when making predictions. The output is attention weights, typically viewed as a heatmap. An application example of this method would be in machine translation, where it shows which words the model is attending to in the source sentence when translating each word in the target sentence.

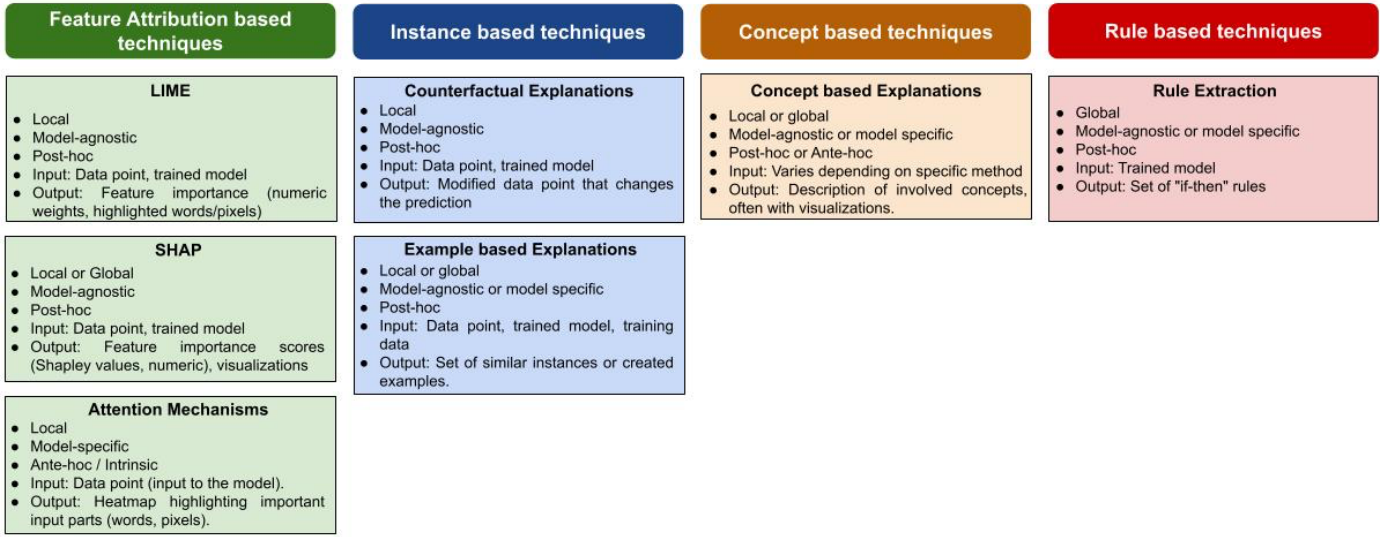

Figure 1: Summarizing and classifying common eXplainable AI (XAI) techniques.

*2) Instance based (or example based) techniques:* These methods show examples (either real or synthetic) to illustrate the model's behavior. They explain "by analogy" or by showing "what-if" scenarios. Common techniques include:
- Counterfactual Explanations [19]: Counterfactual explanations are a post-hoc, generally model-agnostic method for local explanations. They show how the model would make different predictions if certain input features differed, giving "what-if" type responses. They show the least modification to the input features, adequate to change the model's prediction to a specified alternative. Input is a data point and the model; output is a transformed data point (same structure as input) that would result in a different prediction. Mothilal, Ramaravind K., et al. (2020) proposed a framework [4] for generating and evaluating diverse counterfactual explanations based on determinantal point processes. Extending the loan application example, a counterfactual explanation could indicate that a $10,000 boost in the applicant's income would suffice for loan approval.
- Example-Based Explanations [10]: These are often post-hoc and model-agnostic, although a few models like k-nearest neighbor (k-NN) are inherently example-based. These methods find similar cases in the training set or produce counterfactual cases. The input is the trained model. The output is a set of examples to clarify the internal representation of data. The scope can be local (explaining a single prediction) or global (representing the whole model).

*3) Concept based techniques:* These techniques go beyond simple feature importance or examples and tries to explain in terms of abstract concepts that the model has (implicitly or explicitly) learned.

- Concept-Based Explanations [12]: These are typically model-specific and post-hoc explanations and provide a global explanation of the model. They try to discover higher-level concepts that are driving the model's decisions. They work by discovering sets of inputs activating portions of a model, developing a concept, and measuring each discovered concept's contribution to the model's prediction. The input is usually the trained model, and the output describes the concepts involved, usually with visualizations. An example of this type of explanation would be determining that the concept "striped" in an image dataset considerably influences the classification of images into "zebra". Yeh, Chih-Kuan, et al. (2020) investigated concept-based explainability for Deep Neural Networks (DNNs) by defining completeness of concepts, proposed a method to discover interpretable and complete concepts, and introduced an approach to quantify concept importance [13].

*4) Rule based techniques:* This method explicitly generates rules for the model's logic.

- Rule Extraction [30]: These techniques attempt to extract human-readable rules from a trained model. These rules state the model's decision-making process in an "if-then" form. Both post-hoc (when applied to an existing model) and ante-hoc (when the model is to be rule-based from the start) are possible. The scope is typically global. The aim is to create a simplified but interpretable model for the decision process. The input is the trained model, and the output is a collection of rules that approximate the model's decision-making. One of these methods is exemplified by Guido Bologna by showing a rule extraction technique [6] that has been applied to ensembles of decision trees and neural networks.

## C. XAI in Software Engineering

XAI in Software Engineering involves applying XAI techniques to different Software Engineering phases and tasks. A Comprehensive literature study conducted by A. H. Mohammadkhani et. al. (2023) on XAI in Software Engineering presented the following findings [7]:

- Software maintenance: SRL results show this phase is most explored, with 68% of XAI applications in Software Engineering.
- Software development: This phase comprises 16% of XAI applications in software engineering.
- Software management and requirements: These two tasks received less attention, but as reported in the studies surveyed, each accounts for 8% of XAI applications.
- Other tasks: Software design and testing have not been researched.

## III. AI APPLICATIONS, XAI CHALLENGES & XAI METHODS IN EACH SDLC PHASE

In this section, we will go over different phases of the Software Development Lifecycle (SDLC). For each phase, we will cover 1) the applications of AI in each phase, 2) the AI explainability-related challenges in each phase, and 3) XAI techniques tailored to each phase that could help address those challenges. The SDLC is iterative; the stages often repeat multiple times as the software evolves. This paper covers the following SDLC phases:

- Requirement Elicitation: This phase involves discovering the stakeholders' needs and constraints and establishing what the software needs to do.
- Design: The design phase takes the gathered requirements and turns them into a blueprint for the software. It involves creating the system architecture, specifying data structures, algorithms, interfaces, and modules, and outlining how the system will satisfy the requirements.
- Development (Implementation/Coding): This involves coding and implementing the software according to the design specifications. Developers transform the design into a functional software product.
- Testing: In this phase, the software is tested to reveal defects and ensure it functions according to the specifications and requirements. There are several levels of testing, including unit, integration, system, and acceptance testing.
- Deployment and Monitoring: Deployment means making the software available to use by releasing it to users or deploying it into the production environment, and monitoring means continuously tracking the performance of the deployed system and collecting user feedback.
- Maintenance and Evolution: This is the ongoing process of modifying the software after it has been deployed to correct faults, improve performance, or adapt to the changing environment. It includes debugging, new feature additions, and system updates.

## A. Requirement Elicitation Phase

*1) Most common AI Applications:* A literature review from 2023 conducted by Cheligeer C et. al. demonstrated how AI offers significant potential for automating and enhancing requirements elicitation based on the analysis of existing documents to infer key information and even create an initial draft of requirements [2]. Natural Language Processing (NLP) and Large Language Models (LLMs) can be utilized to understand user requirements, identify contradictions, and prioritize requirements [8],[28]. LLMs can also be used to create and improve requirement specifications [15]. AI can also assist in requirements elicitation by generating requirements from high-level user inputs or documents, eliminating ambiguity and identifying missing information [7].

*2) Most common AI Explainability (XAI) Challenges:* XAI challenges often emerge during requirement elicitation, where appropriate XAI techniques can provide support:

- Ambiguity and Incompleteness: Natural language is inherently ambiguous, and requirements are often incomplete. AI may misinterpret requirements or miss crucial details. AI also might misunderstand ambiguous or incomplete user statements, leading to incorrect requirements. [28].
- Tacit Knowledge: Much knowledge about requirements is tacit. Stakeholders have implicit knowledge that they do not even realize they possess or cannot easily articulate. AI struggles with this during requirement elicitation because it needs concrete data [5].
- Bias and Fairness: Training data may contain biases, leading to unfair requirements [7],[9].

*3) Most common & effective XAI techniques:* As per the literature review, the most effective XAI techniques in addressing XAI challenges include:
- LIME/SHAP can help by identifying which features of existing systems [9] or user actions are most influential in producing specific outcomes (e.g., user satisfaction, task completion) [19]; such methods can even bring out latent requirements. Suppose a feature repeatedly has a high SHAP value for positive predictions. In that case, it indicates that this feature is significant to users [12], even if they did not include it as a requirement. SHAP values specifically could assist in detecting biases by measuring the contribution of each feature to the prediction, making it easier to identify if protected attributes (such as race or gender, etc.) or proxies for protected attributes are influencing the requirements. LIME can also be used, but SHAP is generally preferred for its stronger theoretical foundations. For instance, analyzing user interactions with a mobile app prototype using LIME/SHAP could reveal that users who complete a key task often utilize a particular gesture or navigation sequence. This could reveal an unspoken requirement for that gesture or navigation flow, even though users did not express it clearly through interviews or user studies.
- Counterfactual Explanations show how small changes to input features affect the outcome. They can help stakeholders understand the system's sensitivity and identify potential trade-offs [9],[12]. This can be particularly useful for clarifying ambiguous or vague and evolving requirements. For instance, "If we add a requirement for X, how will that affect the system's ability to satisfy Y?" or "If we relax requirement Z, what other requirements become feasible?"

*B. Design Phase*

*1) Most common AI Applications:* AI is used during the design phase of software development for several purposes, including architecture recommendation, design pattern selection, User Interface (UI) & User Experience (UX) design, model selection, and code generation [8]. Specifically, AI can assist in suggesting suitable architectures based on requirements and constraints, suggesting proper design patterns, designing user interface mockups, and even recommending optimal ML models for specific tasks within the system [3],[7].

*2) Most common AI Explainability (XAI) Challenges:* The set of explainability challenges that arise during the design phase are mostly justifying why certain architecture or design pattern is recommended [3]. This includes highlighting the trade-offs that AI considers, such as performance vs. security, etc. Also, there might be areas where AI might be uncertain while generating software design recommendations.

*3) Most common & effective XAI Techniques:* As per the literature review, the most effective XAI techniques in addressing XAI challenges include:
- Counterfactual Explanations: This is a firm fit for both justification and trade-off analysis [5],[33],[35]. Counterfactuals directly answer the question, "What would need to change in the input (requirements, constraints) to get a different design recommendation?" This makes them inherently good at showing trade-offs.
    - Example: "The system recommended a microservices architecture because the requirements emphasized scalability. If high performance was prioritized instead, a monolithic architecture might have been recommended."
- Rule Extraction: If the underlying AI model making the design recommendations is a decision tree or random forest, rule extraction is highly suitable [7], [19]. The rules directly show the decision-making logic.
    - Example: "IF requirement_scalability = HIGH AND requirement_maintainability = MEDIUM THEN architecture = MICROSERVICES." The rules are easy to understand and directly show the trade-offs and the factors influencing the decision.
- Concept-Based Explanations: If the AI can be trained to recognize and reason about high-level concepts (e.g., "scalability," "security," "maintainability"), then concept-based explanations could be very effective [12], [19]. However, this requires defining and identifying relevant concepts, which can be challenging.
    - Example: "The system chose a microservices architecture because of its focus on the concept of *scalability*."

*C. Development Phase*

*1) Most common AI Applications:* In the development (or implementation/coding) phase of the SDLC, AI is primarily used for code generation, code completion, code summarization, and bug detection/repair [26]. AI can also be used for code translation and refactoring. Essentially, AI streamlines the coding process and assists developers by automating repetitive tasks, suggesting code snippets, and identifying potential errors [1].

*2) Most common AI Explainability (XAI) Challenges:* XAI-related challenges in this phase center mostly around correctness and reliability. This includes understanding why the AI made certain suggestions, so that developers can ensure the generated code is functionally correct, secure, and maintainable [23],[34].

*3) Most Common & effective XAI Techniques:* As per the literature review, the most effective XAI techniques in addressing XAI challenges include:
- LIME/SHAP (Feature Attribution): These model-agnostic techniques [18-19] can be very helpful for debugging and understanding specific code suggestions. They highlight which parts of the input (e.g., the natural language prompt, the surrounding code context) were most influential in generating a particular line or block of code. This can help developers understand *why* the AI made a specific suggestion. For instance, if a generated function is incorrect, LIME/SHAP could show that a particular keyword overly influenced the AI in the prompt, or that the AI ignored a crucial part of the surrounding code.
- Example based explanations and counterfactuals: These techniques are very well-suited for addressing correctness and justification [9], [33].
  - Examples: Showing similar, correct code snippets from the training data can help developers understand the learned patterns by the AI and feel confident about the suggestions.
- Counterfactuals: These are particularly effective for justification. They answer the question, "What would have to vary in the input to get a different output?" This can help developers understand the generated code's sensitivity to changes in the requirements or context. For example, "If you remove the requirement for thread safety, the generated code would lack the lock mechanism."

*D. Testing*

*1) Most common AI Applications:* AI in software development testing can be utilized in numerous ways, from test case generation to test prioritization, test oracle generation, fault localization, and even metamorphic testing. LLMs alone show promise in test case generation, oracle generation, and understanding existing tests [15].

*2) Most common AI Explainability (XAI) Challenges:* In the testing phase of software development, the most common XAI challenge is understanding why a particular test case failed. Determining the expected output of a testcase (the "oracle") is often difficult [3].

*3) Most common & effective XAI Techniques*: During the literature review, the most effective XAI techniques in addressing XAI challenges were found to be:
- LIME/SHAP (Feature Attribution) [18-19]: They show which input features (parts of the test case, code being tested, or execution context) were most influential in leading to the model's prediction (pass/fail, specific output, etc.). This helps understand why a test case failed or produced a particular output. For instance, if a test case fails, LIME/SHAP could highlight that a specific input value or a particular line of code was the primary driver of the failure.
- Counterfactuals: These are particularly powerful for justification [9][12]. They answer the question, "What would need to change in the input to get a different output?" In the testing context, this could mean, "What small change to the test case would cause it to pass (or fail)?".

*E. Deployment & Monitoring*

*1) Most common AI Applications*: During the deployment and monitoring stage of SDLC, AI can assist with activities such as anomaly detection, failure prediction, performance analysis, and facilitating continuous integration and delivery. Generative AI (GenAI) can be used for monitoring against possible deployment anomalies and facilitating rollback strategies when needed, streamlining the release process [26]. AI can be used for performance analysis by analyzing metrics and making suggestions for enhancing deployed software products. Measuring and collecting performance metrics is the primary use of GenAI, providing suggestions for improvement. [20].

*2) Most common AI Explainability (XAI) Challenges:* AI applications have several explainability challenges in the SDLC's deployment and monitoring phase. If AI flags a performance anomaly, what is the cause? Is it a genuine problem, or a false alarm? Alternatively, understanding why the AI flagged a particular log entry as suspicious. Why is AI predicting a failure? What are the contributing factors? At the same time, if AI is used for resource management and optimization, it is important to know why AI made a particular scaling, balancing, or allocation decision.

*3) Most common & effective XAI Techniques*: During the literature review, the most effective XAI techniques in addressing XAI challenges were found to be:
- LIME/SHAP (Feature Attribution): These can highlight which performance metrics (CPU usage, memory, latency, etc.) were most influential in triggering the anomaly flag. This helps pinpoint the source of the problem. If the AI is trained on log data, these can show which words or phrases in the log entry were most important for the "suspicious" classification. This helps a human understand why it was flagged. LIME/SHAP can reveal which input features (e.g., system state, recent user actions, etc.) contributed most to the predicted failure. This helps understand the causes and potentially prevent the failure. LIME/SHAP can show which factors (e.g., current/predicted load, resource availability, etc.) drove the scaling/balancing/allocation decision. This provides transparency and allows for auditing.
- Counterfactuals: Extremely valuable for understanding sensitivity and providing actionable insights.
  - Explaining Performance Anomaly: "If the request rate had been 20% lower, the anomaly would not have been flagged."
  - Explaining Suspicious Log Entry: "If the log entry had not contained the phrase 'access denied', it would not have been flagged."

o Explaining Resource Management: "If the predicted load were 10% lower, fewer servers would have been allocated."

*F. Maintenance & Evolution*

*1) Most common AI Applications*: AI can greatly assist during this phase, particularly through Large Language Models (LLMs). AI can help identify and predict potential bugs or vulnerabilities and even suggest or assist in developing a fix. AI can create concise descriptions of what the code does to make understanding and maintenance easy. AI can suggest improvements in the structure and maintainability of the code. AI can be further used to automatically produce or update documentation to keep pace with code changes.

*2) Most common AI Explainability (XAI) Challenges:* Even if an AI suggests a bug fix or a refactoring, developers need to trust that the suggestion is correct and will not introduce new problems. This is especially crucial in maintenance, where changes can have cascading effects. Unquestioningly accepting AI-generated changes is risky.

*3) Most common & effective XAI Techniques:* During the literature review, the most effective XAI techniques in addressing XAI challenges were found to be:
- LIME/SHAP: These can pinpoint the code elements that the AI model associates with a bug, helping developers focus their debugging efforts. They can also show the most significant parts of the code in relation to the summary generated, allowing developers to comprehend more precisely why the summary was generated.
- Counterfactual Explanations: These explanations show how the model's prediction would change if some input features differed. They answer "what if" questions. They are highly relevant for Bug Prediction/Fixing, answering questions like "If this line of code were changed, would the bug still be present?"
- Attention Mechanisms: Attention visualization can show which parts of the input code are most relevant to the generated summary, code, or translation.

## IV. DISCUSSION

Explainable AI (XAI) is crucial for Software Engineering to overcome the 'black-box' problem inherent in AI-aided software development [18]. The application of XAI across various phases of Software Engineering promises to increase trust, transparency, and reliability in the AI-based software development process [31]. Although AI techniques have been proven to improve efficiency and decision-making, their non-transparency typically hinders their use [9],[10]. Nevertheless, even with advancements on XAI in software engineering, some gaps remain to be addressed:
- The lack of standardized evaluation metrics for XAI methods in software engineering makes it challenging to compare and assess the effectiveness of different methods. This gap hinders the development of explainable AI-driven software by complicating explanation quality assessment and comparison of XAI methods across software engineering phases [7],[31]
- XAI methods are often ineffective in explaining the behavior of advanced AI models, such as deep neural networks. This shortcoming limits the development of trustworthy AI-driven software systems by making it difficult to understand the reasoning behind the model's decisions and to identify potential biases or errors [3].
- Most XAI methods provide technical explanations that humans cannot easily interpret. This gap hinders the development of maintainable AI-driven software systems because it becomes difficult for software engineers to understand the AI system's behavior and to make changes or updates accordingly [31].
- XAI methods often lack integration with software development processes, limiting practical use and hindering the development of explainable, trustworthy, and maintainable AI-driven systems. [32].

While existing research findings show that Software Maintenance has received the most attention in XAI research [22], this paper provides a comprehensive overview of XAI techniques across *all* SDLC phases, particularly highlighting opportunities in less-explored areas like requirements and design. However, significant gaps remain in integrating XAI into earlier stages of Software Engineering, such as requirements elicitation and design [7]. Explainability can help refine specifications, enhance requirement traceability, and mitigate potential biases early in the development process [28],[3].

One of this paper's key contributions is the comprehensive Software Development lifecycle phase-specific summarization of XAI techniques to improve explainability in Software Engineering processes. This paper presents XAI techniques for some of the key Software Engineering phases. Overall, through this paper, we want to highlight the necessity of XAI infusion across the Software Development Lifecycle to fill the explainability gap that would promote responsible AI-aided software development [7],[31].

## V. CONCLUSION

Transparency and trustworthiness are functionally imperative in AI systems, particularly within Software Engineering activities [7]. As AI continues to revolutionize various domains, advancing XAI paradigms that elucidate these systems' decision-making processes is critical [35]. Examining XAI techniques in the different phases of Software Engineering offers a fertile ground for addressing the inherent complexity in this field [7].

Enabling explainability is vital for alleviating concerns around reliability and ethical implications in AI systems, as evidenced by numerous studies highlighting the necessity of comprehensible outputs for successful adoption of AI within Software Engineering [32],[2].

The development of robust XAI tools capable of revealing the decision-making processes of AI models will enhance user confidence and facilitate broader acceptance of AI solutions across diverse fields [35]. To the best of our knowledge, this is the first work that presents a comprehensive overview of XAI techniques tailored to each phase of the Software Development Life Cycle (SDLC). By doing so, we aim to promote explainable AI in Software Engineering and facilitate the practical use of complex AI models in AI-driven software development.

Future research needs to explore the optimal application of the tested and realized XAI approaches in agile and DevOps focused development paradigms. Also, research must aim to formulate benchmarking structures that enable fair comparison among XAI approaches to ensure a better fit with requirements in real-world applications. This is essential in software development for developing standard evaluation criteria for XAI in Software Engineering to assess effectiveness uniformly across all approaches [12].

## REFERENCES

[1] I. Baskhad and S. Tim, "Program Code Generation with Generative AIs," Algorithms, vol. 17, no. 2, 2024, doi: 10.3390/a17020062.

[2] Cheligeer C, et. al, Machine learning in requirements elicitation: a literature review. *Artificial Intelligence for Engineering Design, Analysis and Manufacturing*. 2022;36:e32. doi:10.1017/S0890060422000166

[3] S. Martínez-Fernández et al., "Software Engineering for AI-Based Systems: A Survey," ACM Transactions on Software Engineering and Methodology, vol. 31, no. 2, pp. 1–59, Apr. 2022, doi: 10.1145/3487043.

[4] Mothilal, Ramaravind K., et al. "Explaining Machine Learning Classifiers through Diverse Counterfactual Explanations." *Proceedings of the 2020 Conference on Fairness, Accountability, and Transparency*, ACM, 2020, pp. 607–17. *DOI.org (Crossref)*, https://doi.org/10.1145/3351095.3372850.

[5] L. Chazette, et. al, "Explainable software systems: from requirements analysis to system evaluation," Requirements Engineering, 2022, doi: 10.1007/s00766-022-00393-5.

[6] Bologna, Guido. "A Rule Extraction Technique Applied to Ensembles of Neural Networks, Random Forests, and Gradient-Boosted Trees." *Algorithms*, vol. 14, no. 12, Nov. 2021, p. 339. *DOI.org (Crossref)*, https://doi.org/10.3390/a14120339.

[7] A. H. Mohammadkhani et. al, "A Systematic Literature Review of Explainable AI for Software Engineering," arXiv.org, Feb. 2023, doi: 10.48550/arxiv.2302.06065.

[8] Hutchinson, Ben, et al. "Towards Accountability for Machine Learning Datasets: Practices from Software Engineering and Infrastructure." Proceedings of the 2021 ACM Conference on Fairness, Accountability, and Transparency, ACM, 2021, pp. 560–75. DOI.org (Crossref), https://doi.org/10.1145/3442188.3445918.

[9] A. Bennetot et al., "A Practical tutorial on Explainable AI Techniques," ACM Computing Surveys, 2024, doi: 10.1145/3670685.

[10] T. Clement, et al, "XAIR: A Systematic Metareview of Explainable AI (XAI) Aligned to the Software Development Process," Machine Learning and Knowledge Extraction, vol. 5, no. 1, pp. 78–108, Jan. 2023, doi: 10.3390/make5010006.

[11] K. Peter, "The Use of AI in Software Engineering: A Synthetic Knowledge Synthesis of the Recent Research Literature," Information, vol. 15, no. 6, 2024, doi: 10.3390/info15060354.

[12] S. Ali et al., "Explainable Artificial Intelligence (XAI): What we know and what is left to attain Trustworthy Artificial Intelligence," Information Fusion, 2023, doi: 10.1016/j.inffus.2023.101805.

[13] Yeh, Chih-Kuan, et al. "On completeness-aware concept-based explanations in deep neural networks." *Advances in neural information processing systems* 33 (2020): 20554-20565.

[14] P. J. Phillips et al., "Four Principles of Explainable Artificial Intelligence," null, 2021, doi: 10.6028/nist.ir.8312.

[15] Hou, Xinyi, et al. "Large Language Models for Software Engineering: A Systematic Literature Review." *ACM Transactions on Software Engineering and Methodology*, vol. 33, no. 8, Nov. 2024, pp. 1–79. *DOI.org (Crossref)*, https://doi.org/10.1145/3695988.

[16] Samek, Wojciech, et al. "Evaluating the Visualization of What a Deep Neural Network Has Learned." *IEEE Transactions on Neural Networks and Learning Systems*, vol. 28, no. 11, Nov. 2017, pp. 2660–73. *IEEE Xplore*, https://doi.org/10.1109/TNNLS.2016.2599820

[17] Messalas, Andreas, et al. "Model-Agnostic Interpretability with Shapley Values." *2019 10th International Conference on Information, Intelligence, Systems and Applications (IISA)*, IEEE, 2019, pp. 1–7. *DOI.org (Crossref)*, https://doi.org/10.1109/IISA.2019.8900669

[18] D. Minh, H. X. Wang, Y. F. Li, and T. N. Nguyen, "Explainable artificial intelligence: a comprehensive review," Artificial Intelligence Review, pp. 1–66, Nov. 2021, doi: 10.1007/s10462-021-10088-y.

[19] Mersha, Melkamu, et al. "Explainable Artificial Intelligence: A Survey of Needs, Techniques, Applications, and Future Direction." *Neurocomputing*, vol. 599, Sept. 2024, p. 128111. *DOI.org (Crossref)*, https://doi.org/10.1016/j.neucom.2024.128111.

[20] Simaremare, Mario, et al. "Exploring the Potential of Generative AI: Use Cases in Software Startups." *Agile Processes in Software Engineering and Extreme Programming – Workshops*, edited by Lodovica Marchesi et al., vol. 524, Springer Nature Switzerland, 2025, pp. 3–11. *DOI.org (Crossref)*, https://doi.org/10.1007/978-3-031-72781-8_1.

[21] M. Merkel and J. Dorpinghaus, "A case study on the transformative potential of AI in software engineering on LeetCode and ChatGPT," 2025, doi: https://doi.org/10.48550/arXiv.2501.03639.

[22] Görkem Giray. 2021. A software engineering perspective on engineering machine learning systems: State of the art and challenges. J. Syst. Softw. 180, C (Oct 2021). https://doi.org/10.1016/j.jss.2021.111031

[23] V. Hassija et al., "Interpreting Black-Box Models: A Review on Explainable Artificial Intelligence," Cognitive Computation, vol. 16, pp. 45–74, Aug. 2023, doi: 10.1007/s12559-023-10179-8.

[24] A. Nguyen-Duc et al., "Generative Artificial Intelligence for Software Engineering -- A Research Agenda," arXiv.org, Oct. 2023, doi: 10.48550/arxiv.2310.18648..

[25] F. A. Batarseh, R. Mohod, A. Kumar, and J. C. Bui, "The application of artificial intelligence in software engineering: a review challenging conventional wisdom," Data Democracy, pp. 179–232, Jan. 2020, doi: 10.1016/b978-0-12-818366-3.00010-1.

[26] D. Russo, "Navigating the Complexity of Generative AI Adoption in Software Engineering," ACM Transactions on Software Engineering and Methodology, 2023, doi: 10.1145/3652154.

[27] W. J. von Eschenbach and J. R. Warren, "Transparency and the Black Box Problem: Why We Do Not Trust AI," Philosophy & Technology, pp. 1–16, Sep. 2021, doi: 10.1007/s13347-021-00477-0.

[28] C. Gao, X. Hu, S. Gao, X. Xia, and Z. Jin, "The Current Challenges of Software Engineering in the Era of Large Language Models," ACM Transactions on Software Engineering and Methodology, 2025, doi: 10.1145/3712005.

[29] K. A. Eldrandaly, M. Abdel-Basset, M. Ibrahim, and N. M. Abdel-Aziz, "Explainable and secure artificial intelligence: taxonomy, cases of study, learned lessons, challenges and future directions," Enterprise Information Systems, Jul. 2022, doi: 10.1080/17517575.2022.2098537.

[30] G. Vilone and L. Longo, "Classification of Explainable Artificial Intelligence Methods through Their Output Formats," Machine Learning and Knowledge Extraction, vol. 3, no. 3, pp. 615–661, Aug. 2021, doi: 10.3390/make3030032.

[31] W. Yang et al., "Survey on Explainable AI: From Approaches, Limitations and Applications Aspects," Human-Centric Intelligent Systems, vol. 3, no. 3, pp. 161–188, Aug. 2023, doi: 10.1007/s44230-023-00038-y.

[32] Z. U. Islam, "Software Engineering Methods for Responsible Artificial Intelligence," Adaptive Agents and Multi-Agent Systems, 2021, doi: 10.5555/3463952.3464248.

[33] A. Adadi and M. Berrada, "Peeking Inside the Black-Box: A Survey on Explainable Artificial Intelligence (XAI)," IEEE Access, vol. 6, pp. 52138–52160, Sep. 2018, doi: 10.1109/access.2018.2870052.